\documentclass[a4paper,12pt]{article}
\usepackage{feynmp-auto,expdlist}
\usepackage{amsmath, amsfonts}
\usepackage{graphicx,arydshln}
\usepackage{enumerate}
\usepackage{hyperref}
\usepackage{latexsym}
\usepackage{dsfont}
\usepackage{hepnicenames}
\usepackage{enumerate}
\usepackage{soul}
\usepackage[normalem]{ulem}
\usepackage{comment}

\newcommand{\footnoten}[1]{}

\usepackage[font={small},textfont={it}]{caption} 

\usepackage{units}

\usepackage{mathrsfs,graphicx,rotating,amsmath,amsfonts,mathtools,booktabs,amssymb,wasysym}
\usepackage{hyperref}\usepackage{slashed}
\usepackage[nosort]{cite}
\usepackage[table,xcdraw,dvipsnames]{xcolor}
\usepackage{bm}
\usepackage{multirow,multicol}
\hypersetup{colorlinks,bookmarksopen,bookmarksnumbered,
	linkcolor=blus,pdfstartview=FitH,urlcolor=rossos,citecolor=verde}
\allowdisplaybreaks

\newcommand{\myfootnote}[1]{}
\newcommand{\myomit}[1]{{\color{gray}#1}}
\renewcommand{\myomit}[1]{}

\renewcommand{\[}{\left[}

\def\Lag{\mathscr{L}}

\newcommand{\mio}[1]{}

\def\bpm{\begin{pmatrix}}
	\def\epm{\end{pmatrix}}

\usepackage{mathrsfs}

\newcommand{\fig}[1]{~\ref{fig:#1}}
\newcommand{\sfrac}[2]{#1/#2}

\allowdisplaybreaks
\usepackage{multicol}
\usepackage{color}
\definecolor{rosso}{cmyk}{0,1,1,0.4}
\definecolor{rossos}{cmyk}{0,1,1,0.55}
\definecolor{rossoc}{cmyk}{0,1,1,0.2}
\definecolor{blu}{cmyk}{1,1,0,0.3}
\definecolor{blus}{cmyk}{1,1,0,0.6}
\definecolor{bluc}{cmyk}{1,1,0,0.1}
\definecolor{verde}{cmyk}{0.92,0,0.59,0.25}
\definecolor{verdec}{cmyk}{0.92,0,0.59,0.15}
\definecolor{verdes}{cmyk}{0.92,0,0.59,0.4}

\newcommand{\eq}[1]{~{\rm (\ref{eq:#1})}}

\newcommand{\GeV}{\,{\rm GeV}}
\newcommand{\TeV}{\,{\rm TeV}}

\newcommand{\pb}{\,{\rm pb}}

\newcommand{\ab}{\,{\rm ab}}

\def\circa#1{\,\raise.3ex\hbox{$#1$\kern-.75em\lower1ex\hbox{$\sim$}}\,}

\newcommand{\beq}{\begin{equation}}
\newcommand{\eeq}{\end{equation}}

\newcommand{\bea}{\begin{eqnarray}}
\newcommand{\eea}{\end{eqnarray}}
\newcommand{\be}{\begin{equation}}
\newcommand{\ee}{\end{equation}}
\font\tenrsfs=rsfs10 at 12pt
\font\sevenrsfs=rsfs7
\font\fiversfs=rsfs5
\newfam\rsfsfam
\textfont\rsfsfam=\tenrsfs
\scriptfont\rsfsfam=\sevenrsfs
\scriptscriptfont\rsfsfam=\fiversfs

\newsavebox\MBox

\renewenvironment{thebibliography}[1]
{\begin{multicols}{2}[\section*{\refname}]%
		\@mkboth{\MakeUppercase\refname}{\MakeUppercase\refname}%
		\list{\@biblabel{\@arabic\c@enumiv}}%
		{\settowidth\labelwidth{\@biblabel{#1}}%
			\leftmargin\labelwidth
			\advance\leftmargin\labelsep
			\@openbib@code
			\usecounter{enumiv}%
			\let\p@enumiv\@empty
			\renewcommand\theenumiv{\@arabic\c@enumiv}}%
		\sloppy
		\clubpenalty4000
		\@clubpenalty \clubpenalty
		\widowpenalty4000%
		\sfcode`\.\@m}
	{\def\@noitemerr
		{\@latex@warning{Empty `thebibliography' environment}}%
		\endlist\end{multicols}}

\renewcommand{\L}\Lag

\def\circa#1{\,\raise.3ex\hbox{$#1$\kern-.75em\lower1ex\hbox{$\sim$}}\,}
\makeatletter

\font\ital=cmu10

\def\hhref#1{\href{http://arxiv.org/abs/#1}{arXiv:#1}}
\usepackage{xstring}
\newcommand{\hhrefq}[1]{\IfSubStr{#1}{:}{\href{http://inspirehep.net/search?ln=en&ln=en&p=#1&of=hb&action_search=Search&sf=&so=d&rm=&rg=25&sc=0}{InSpire:#1}}{\hhref{#1}}}

\def\art{\@ifnextchar[{\eart}{\oart}}
\def\eart[#1]#2#3#4#5#6{{\rm #2}, {\em #3 \bf #4} {\rm (#6) #5} ({\em #1})}
\def\article{\@ifnextchar[{\earticle}{\oarticle}}
\def\oarticle#1#2#3#4#5#6{{\rm #1}, {\ital `#6'}, {\rm #2 #3 (#5) #4}}
\def\earticle[#1]#2#3#4#5#6#7{{\rm #2}, {\ital `#7'}, {\rm #3 #4 (#6) #5}  [\hhrefq{#1}]}
\def\hepart[#1]#2{{\rm #2, \sl#1}}
\def\heparticle[#1]#2#3{#2, {\ital `#3'} [\hhrefq{#1}]}
\newcommand{\doi}[1]{\href{http://dx.doi.org/#1}{[link]}}

\newcommand{\hhrefqq}[1]{\IfBeginWith{#1}{10.}{\href{https://doi.org/#1}{doi:#1}}{\hhrefq{#1}}}
\def\earticle[#1]#2#3#4#5#6#7{{\rm #2}, {\ital `#7'}, {\rm #3 #4 (#6) #5}  [\hhrefqq{#1}]}

\renewenvironment{thebibliography}[1]
{\begin{multicols}{2}[\section*{\refname}]%
		\@mkboth{\MakeUppercase\refname}{\MakeUppercase\refname}%
		\list{\@biblabel{\@arabic\c@enumiv}}%
		{\settowidth\labelwidth{\@biblabel{#1}}%
			\leftmargin\labelwidth
			\advance\leftmargin\labelsep
			\@openbib@code
			\usecounter{enumiv}%
			\let\p@enumiv\@empty
			\renewcommand\theenumiv{\@arabic\c@enumiv}}%
		\sloppy
		\clubpenalty4000
		\@clubpenalty \clubpenalty
		\widowpenalty4000%
		\sfcode`\.\@m}
	{\def\@noitemerr
		{\@latex@warning{Empty `thebibliography' environment}}%
		\endlist\end{multicols}}

\newcounter{alphaequation}[equation]
\def\thealphaequation{\theequation\hbox to
	0.6em{\hfil\alph{alphaequation}\hfil}}
\def\eqnsystem#1{
	\def\@eqnnum{{\rm (\thealphaequation)}}
	\def\@@eqncr{\let\@tempa\relax \ifcase\@eqcnt \def\@tempa{& & &} \or
		\def\@tempa{& &}\or \def\@tempa{&}\fi\@tempa
		\if@eqnsw\@eqnnum\refstepcounter{alphaequation}\fi
		\global\@eqnswtrue\global\@eqcnt=0\cr}
	\refstepcounter{equation} \let\@currentlabel\theequation \def\@tempb{#1}
	\ifx\@tempb\empty\else\label{#1}\fi
	\refstepcounter{alphaequation}
	\let\@currentlabel\thealphaequation
	\global\@eqnswtrue\global\@eqcnt=0 \tabskip\@centering\let\\=\@eqncr
	$$\halign to \displaywidth\bgroup \@eqnsel\hskip\@centering
	$\displaystyle\tabskip\z@{##}$&\global\@eqcnt\@ne
	\hskip2\arraycolsep\hfil${##}$\hfil& \global\@eqcnt\tw@\hskip2\arraycolsep
	$\displaystyle\tabskip\z@{##}$\hfil
	\tabskip\@centering&\llap{##}\tabskip\z@\cr}
\def\endeqnsystem{\@@eqncr\egroup$$\global\@ignoretrue} \makeatother

\definecolor{Gray}{gray}{0.95}

\def\bal#1\eal{\begin{align}#1\end{align}}

\begin{document}
\thispagestyle{empty}
\vspace{0.1cm}
\begin{center}
{\LARGE \bf \color{rossos} A boosted muon collider}\\[2ex]
\vspace{1cm}
{\bf\large Daniele Barducci, Alessandro Strumia}  \\[5mm]
{\em Dipartimento di Fisica, Universit{\`a} di Pisa, Italia}\\[4ex]

{\bf\color{blus}Abstract}
\begin{quote} \large 
A muon collider could produce the heavier Standard Model particles
with a boost, for example in resonant processes such as $\mu^-\mu^+\to h$ or $\mu^-\mu^+\to Z$.
We propose  machine configurations that produce the boost (asymmetric beam energies, tilted beams)
and estimate how much the luminosity is reduced or perhaps enhanced.
The feasibility of the proposed configurations, as well as an estimation of the beam-induced backgrounds and beam energy spread, needs to be evaluated in order to derive more solid conclusions on the physics potential of such boosted collider configurations.
If achievable, the boost can provide new interesting observational opportunities.
For example it can significantly enhance the sensitivity to  long-lived new particles
decaying in a far-away detector, such as dark higgses or sterile neutrinos produced in $h$ or $Z$ decays.
\end{quote}
\end{center}

\setcounter{tocdepth}{1}
\tableofcontents

\newpage
\normalsize

\section{Introduction}
The luminosity of $e^-e^+$ circular colliders, even with 100 km length,
strongly decreases with their energy above 100 GeV~\cite{CEPC,FCCee}.
So $e^- e^+$ colliders maximise the collision energy
by performing symmetric head-on collisions with $\sqrt{s} \simeq 2E_{\rm beam}$.
On the other hand, the luminosity of $\mu^-\mu^+$ circular colliders grows roughly quadratically with their energy~\cite{2103.14043,2203.07261,mu_beam_params}
and would allow to produce the heavier SM particles $Z, h$, etc using a smaller 
radius
$R = E_{\rm beam}/e B \approx 33\,{\rm m}\, (E_{\rm beam}/100\GeV) (10\,{\rm T}/B)$,
where $B$ is the collider magnetic field.

\smallskip

Overall, at $\sqrt{s} = M_{Z}$ an $e^- e^+$ collider is expected to achieve  a larger luminosity than a $\mu^-\mu^+$ collider.
A $\mu^-\mu^+$ collider is nevertheless considered interesting 
at $\sqrt{s}=M_h$ (because the muon Higgs Yukawa coupling allows for $\mu^-\mu^+\to h$ resonant Higgs production~\cite{2203.04324})
and at $\sqrt{s}=2M_t$ (because low luminosity is enough for a precise measurement of the top quark mass~\cite{2203.17197}). However $\mu^-\mu^+$ colliders with energy below the TeV are challenging both from the point of view of the machine and detector design and the high level of background. Moreover, producing an on-shell resonant Higgs requires a beam energy spread comparable to its width over mass ratio, around $ 10^{-5}$. While 
engineering such tiny energy spread might perhaps be possible, it poses a challenge for the accelerator design that needs to be addressed~\cite{2203.04324,2203.07261}.

\medskip

We here explore the possibility that
variations from the optimal symmetric head-on collision geometry could be interesting at a muon collider.

\smallskip

One  technique, used in the PEP-II and KEK$\,$B $e^- e^+ $ colliders at GeV-scale energy, 
employs head-on collisions of two beams with asymmetric energies $E_+$ and $E_-$, such that $s\simeq 4 E_+ E_-$.
This generically allows to produce boosted particles, and was used to resonantly produce $e^- e^+\to\Upsilon$.
A muon collider could resonantly produce one heavy SM particle with given boost,
such as $\mu^-\mu^+\to Z$ or $\mu^-\mu^+\to h$.
In section~\ref{boost} we explore how different $\mu^-\mu^+$ collision geometries can affect the luminosity.
Asymmetric head-on collisions give a mild luminosity loss, while
tilted collisions (discussed in~\cite{2211.05240} for a $e^- e^+$ collider)
could allow to produce boosted heavy SM particles with enhanced luminosity,
if a dedicated non-standard beam optics can be invented. We do not investigate the technical feasibility of these machines configurations and the related interaction regions, which need to be evaluated by accelerator and experimental experts,
concentrating our study on identifying interesting physics cases that can take advantage of these non-standard collider options.

\medskip

In particular, we show how the production of boosted heavy SM particles 
offers a significantly enhanced sensitivity to searches for long-lived weakly interacting new particles,
as the boosted kinematics allows to concentrate them towards a far-away detector,
that can only cover a small solid angle.
Section~\ref{h} studies the case of resonant $\mu^-\mu^+\to h$ Higgs production,
section~\ref{Z} studies $\mu^-\mu^+\to Z$,
section~\ref{NR} considers generic non-resonant processes.

\medskip

Various experimental effects must be thoroughly studied in order to assess the robustness of our results. Beam-induced background effects, that might have a significant impact in the forward-direction even if the detector is positioned far-away from the interaction point, must be investigated, together with beam-energy spread effects that might deplete the resonant cross-section for $h,Z$ production.
\medskip

Conclusions are given in section~\ref{concl}.

\section{Boosting collisions at a muon collider}\label{boost}
We here  discuss the luminosity of two beams of particles with energies $E_+$ and $E_-$,
masses $m_+$ and $m_-$ that collide with relative angle $\theta$
between their spatial momenta $\vec p_+$ and $\vec p_-$, such that the
collision energy  is
\beq\label{eq:s}
s \equiv (p_+^\mu+p_-^\mu)^2  = 2 \left(E_+ E_-  - |\vec p_+| |\vec p_-| \cos\theta +\frac{m_+^2+m_-^2}{2}\right) 
\eeq
where $p_\pm^\mu = (E_\pm, \vec p_\pm)$ are the quadri-momenta.

\subsection{Head-on collisions with asymmetric beam energies}\label{headon}
The collision energy $s$ is maximal in the usual case of head-on collisions, corresponding to $\theta=\pi$.
We consider two $\mu^\pm$ beams with energies $E_\pm= \gamma_\pm m_\mu$
circulating in rings with radii $R_\pm$,
and thereby with magnetic fields $B_{\pm}=E_{\pm}/R_\pm e$, 
containing $N_{b\pm}$ bunches of 
$N_{\pm}$ muons $\mu^\pm$ each.
In order to collide, the bunches must be equi-spaced, $R_+/N_{b+} = R_-/N_{b-}$.
We denote as $f$ the repetition rate of acceleration cycles.
The instantaneous luminosity
can be straightforwardly computed by adapting the standard computation of, e.g.,~\cite{2303.08533} to the general case. By considering the fact that the beams have an asymmetric configuration one obtains
\be
{\cal L} = \frac{f N_+ N_-}{2\pi(\sigma_{T+}^2+\sigma_{T-}^2)} \sum_{j=0}^{\infty}{\rm exp}(- \frac{2\pi R_+}{\gamma_+ \tau_\mu N_{b+}}j){\rm exp}(- \frac{2\pi R_-}{\gamma_- \tau_\mu N_{b-}}j) \ .
\ee 
By summing the geometric series and expanding the exponential one thus arrives at 
\beq \label{eq:Lheadon}
{\cal L}=\frac{\tau_\mu}{R_+/\gamma_+ N_{b+}+R_-/\gamma_- N_{b-}} \frac{f N_+ N_-}{(2\pi)^2 (\sigma_{T+}^2 + \sigma_{T-}^2)} \ ,
\eeq
where $\tau_\mu$ is the $\mu$ lifetime and $\sigma_{T\pm}$ are the bunches transverse sizes. This expression reduces to the usual luminosity 
for beams of unstable particles in the symmetric case~\cite{2303.08533}.
Head-on  symmetric collisions with $E_+ \neq E_- $ have
suppressed instanteneous luminosity compared to the symmetric case $E_+=E_-$ with the same $s$,
because of two factors:
\begin{enumerate}
\item the first term of eq.\eq{Lheadon}
implies that the muons with lower energy (say $E_-$) 
decay faster than those with higher energy, 
not allowing to use the more boosted life-time to gain luminosity;
\item the transverse bunch sizes $\sigma_{T\pm}$ in the second term of eq.\eq{Lheadon} are given by
\beq
\frac{1}{\sigma_{T\pm}^2} = \frac{E_\pm^2 f_{\rm hg} \sigma_{\delta\pm}}{m_\mu \epsilon_{L\pm}\epsilon_{T\pm}} \ ,
\eeq
where $\sigma_{\delta\pm}$ is the fractional beam energy spread,
$f_{\rm hg}\approx 0.76$ is the hourglass factor~\cite{Herr:2003em,2303.08533} that limits the maximal focusing achievable,
and $\epsilon_{L,T}$ are the longitudinal and transverse emittances of the bunches,
roughly conserved during the acceleration process.
This means that a higher-energy beam can be better focused,
but this gain is lost when colliding it with a thicker lower energy beam
(say $\sigma_{T-}\gg \sigma_{T+}$).
\end{enumerate}
As a result, by using eq.~\ref{eq:Lheadon}, the luminosity at fixed $s$ gets reduced as 
\beq \label{eq:RL}
\frac{{\cal L}(E_+ \neq E_-)}{{\cal L}(E_+=E_-)} = \frac{2E_+ E_- }{E_+^2+ E_-^2} =\frac{1}{2\gamma^2-1}\eeq
in the limit of equal quality beams, $\epsilon_{L+}=\epsilon_{L-}$,
$\epsilon_{T+}=\epsilon_{T-}$, $\sigma_{\delta+}=\sigma_{\delta-}$.
The luminosity loss can be mitigated if, e.g., the lower energy beam has larger energy spread (say $\sigma_{\delta-}>\sigma_{\delta+}$).
The latter term in eq.\eq{RL} shows the result as function of the boost factor $\gamma\ge 1$
of a resonantly produced particle $\mu^-\mu^+\to X$.
As discussed in the next sections,
the boost $\gamma$ enhances the sensitivity of specific searches,
partially compensating for this luminosity loss.

\subsection{Rear-end collisions with asymmetric beam energy}
Rear-end collisions among beams circulating in the same direction 
(corresponding to $\theta=0$)
allow to reduce $\sqrt{s}$ while keeping comparable beam energies.
This collision geometry has two problems.
First, $s\simeq m_\mu^2[2+ (E_+^2+E_-^2)/E_+ E_-]$ is too low
if one wants to produce the heavier SM particles
rather than GeV-scale particles.
Second, the collision region is too long, $\sigma_L/\delta v $,
where $\sigma_L \sim   {\rm mm}$  is a realistic beam length~\cite{2203.07261} and
$\delta v \simeq (1/\gamma_-^2-1/\gamma_+^2)/2 $
for $ \gamma_\pm \gg 1$ is the small relative velocity.

\begin{figure}[t]
\begin{center}
$$\includegraphics[width=0.48\textwidth]{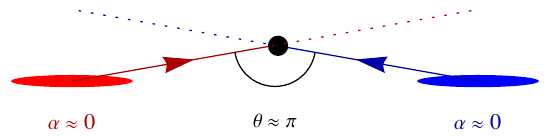} \qquad\includegraphics[width=0.23\textwidth]{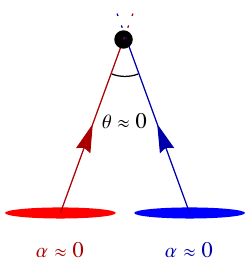}$$
\caption{\label{fig:beams} {\bfseries Left}: the usual  nearly-head-on collision geometry,
where the luminosity is increased by tilting the bunches by a small angle.
{\bfseries Right}: the collision geometry that could enhance the luminosity of processes such as $\mu^+\mu^- \to h$ with $E\gg M_h$.} 
\end{center}
\end{figure}

\subsection{Oblique collisions with equal beam energy}
In order to reduce the collision energy of eq.\eq{s} by the desired amount compared to the beam energies
we consider oblique collisions with generic angle $\theta$.
For $E_\pm \gg\sqrt{s}$ one needs $\theta \ll 1$, and eq.\eq{s}  reduces to
$s \simeq E_+ E_- \theta^2$, where $\theta$ is expressed in radians.
Collisions are then dominantly due to the transverse relative velocity between beams.
We here compute the luminosity for generic $\theta$.
Assuming Gaussian bunches with equal longitudinal bunch sizes $\sigma_L$
and equal transverse beam sizes $\sigma_T$ implies that the tilt angle affects the luminosity as~\cite{Suzuki:1976xe}
\beq  \label{eq:LumiTilt}
\frac{ {\cal L}(\theta) }{{\cal L}(\pi) }=  
\left[1+\frac{\sigma_L^2}{\sigma_T^2}\cot^2{\frac{\theta}{2}} \right]^{-1/2}
 \eeq
where ${\cal L}(\pi)$ corresponds to maximal $s$.
Bunches are usually focused such that $\sigma_{T}\ll \sigma_L$ at the collision point
(planned values are $\sigma_{T}\sim \mu{\rm m}$ and $\sigma_L \sim {\rm mm}$~\cite{2203.07261}),
so even a small deviation from head-on $|\theta-\pi| \circa{>} \sigma_{T}/\sigma_L$ causes a significant
drop in luminosity.
This can be avoided by colliding in a crab-like configuration~\cite{Oide:1989qz,Sun:2009zzj}
by tilting the bunches in order to recover the missing geometric overlap. We define the crabbing angle $\alpha$ as the angle between the longitudinal size of the bunches $\sigma_L$ and the direction identified by the difference of the bunches spacial momenta. The zero-tilting case of eq.\eq{LumiTilt} (with $\sigma_L$ along the beam axis) corresponds to $\alpha=\sfrac{(\pi-\theta)}{2}$.
The optimal collision geometry, with bunches rotated in the way illustrated in both panels of fig.\fig{beams}, is obtained for $\alpha=0$:
\beq  \label{eq:LumiTiltCrab}
 \frac{ {\cal L}(\theta ,\alpha) }{{\cal L}(\pi, 0) }
=\sqrt{\frac{1-\cos\theta}{\displaystyle 1+\frac{\sigma_L^2}{\sigma_T^2} - \left(1-\frac{\sigma_L^2}{\sigma_T^2} \right)\cos\alpha}} =
\sin\frac{\theta}{2}\qquad\hbox{for $\alpha=0$} 
\eeq
A muon collider collision with reduced $s$ 
(for example $\sqrt{s}=M_h$ when considering $\mu^-\mu^+\to h$)
can be obtained from beams with equal energy
$E_\pm  = E \gg m_\mu = m_{\pm}$ 
choosing the collision angle $\theta$ as
\be\label{eq:onshell_angle}
\theta = \arccos\left(1- \frac{s}{2 E^2} \right)  ,
\ee
so that $\sqrt{s}\ll E$ is obtained from a small $\theta \simeq \sqrt{s}/E$, corresponding to quasi-parallel beams.
In this limit eq.\eq{LumiTilt} reduces to
${ {\cal L}(\theta) }/{{\cal L}(\pi) }\simeq (\theta/2)( \sigma_T/\sigma_L)$:
the double suppression by
$\theta\ll1$ and by $\sigma_T/\sigma_L \ll 1$ implies a big luminosity loss with the usual
geometry of beams.
The latter suppression is avoided colliding tilted bunches with the optimal angle $\alpha=0$,
such that eq.\eq{LumiTiltCrab} reduces to 
\beq  \label{eq:LumiTiltCrabalpha}
 \frac{ {\cal L}(\theta ,\alpha) }{{\cal L}(\pi, 0) }\simeq\frac{ \theta}{2}\simeq \frac{1}{\gamma}\ .\eeq 
This suppression in the luminosity is milder
than the enhancement ${\cal L}(\pi, 0)  \propto E^2$ 
of the luminosity of a $\mu^-\mu^+$ circular collider as its beam energy $E$ is increased.
So a $\mu^-\mu^+$ circular collider with nearly-parallel beams of energy $E$
could potentially produce heavier SM particles of mass $\sqrt{s}=M$ with luminosity 
enhanced by $\gamma=2E/M$ compared to usual head-on collisions with $E=M/2$.\footnote{Since a muon collider matches the scaling $\sigma\propto 1/E$ of bunch size with particle wave-length, the scaling factor $\gamma$
can be also obtained from the simpler problem of colliding two particles: 
$\gamma$ arises as the factor $t_{\rm cm} = \gamma t_{\rm lab}$ 
that relates the collision time in the laboratory and center-of-mass frames.}
However this would need a dedicated machine optics with the beam geometry illustrated
in the right panel of fig.\fig{beams} to achieve $\alpha \approx 0$.
It could be interesting to explore if this can be realistically realized.
A pre-collision region with a time-dependent magnetic field in a size $2R'$ could rotate conventional head-on beams 
by nearly $\pm 90^\circ$ without rotating the bunches, providing collisions with $\sqrt{s}=e B R'$
(e.g.\ $\sqrt{s}=M_h$  $R'=M_h/eB = 42\,{\rm m}$ for $B=10\,{\rm T}$),
but the real difficulty is achieving focus at the collision point.
With this geometry the beam energy spread $\sigma_\delta$
negligibly contributes to the spread in $\sqrt{s}$,
that is produced in the rotation process.
We do not explore if/how these wild speculations could be realistically implemented.

\smallskip

Even if this luminosity gain cannot be practically achieved,
the boosted produced particles lead to the extra gain in the sensitivity 
of some specific searches discussed in the next sections.
So even the simpler option of limiting the luminosity loss as in section~\ref{headon} could be interesting. We thus explore in the following sections the physics potential of such collider configurations with beam energies ranging from $E_\pm = E_{h,Z}/2$ up to $E_\pm = 5\;$TeV.

\section{Resonant $\mu^-\mu^+\to h$ production}\label{h}
We consider $\mu^-\mu^+ \to h$ resonant production of Higgs bosons  with boost
$\gamma_h= E_h/M_h$.
 As discussed in section~\ref{boost}, this could be done at a muon collider
with luminosity possibly reduced by $\sim 1/\gamma_h^2$ or perhaps enhanced by $\gamma_h$.
As mentioned in the Introduction, producing an on-shell resonant Higgs requires a small beam energy spread $\sim \Gamma_h/M_h \sim 10^{-5}$, which poses a challenge for the accelerator design~\cite{2203.07261} but that, if overcome, can lead to interesting physics possibilities~\cite{2203.04324}. Around the Higgs peak one expects
\beq
\sigma(\mu^-\mu ^+\to h \to X) = \frac{4\pi \Gamma_h^2\, {\rm BR}(h\to \mu^-\mu^+) }{(s-M_h^2)^2 + M_h^2 \Gamma_h^2}
{\rm BR}(h \to X)\ .\eeq
Here $\Gamma_h/M_h \approx 3\,\times 10^{-5}$ is the narrow Higgs width,
and ${\rm BR}(h\to \mu^-\mu^+)\approx 0.22\times10^{-3}$ in the SM.
Then the peak cross section would be $\sigma (\mu^-\mu^+\to h)=4\pi\, \hbox{BR}(h\to\mu\mu)/M_h^2 \approx 71\,{\rm pb}$.
Initial state radiation reduces it down to $\sigma \approx 37\,{\rm pb}$,
and the unknown energy beam spread could further reduce it down to
$\sigma \approx 22\,{\rm pb}$~\cite{1607.03210}.
We here assume this latter value.
This is comparable to the Higgs production cross section at a $pp$ collider,
$\sigma(pp\to h)\sim 50\,{\rm pb}$ at $\sqrt{s}=13\TeV$, dominated by $gg\to h$.

\smallskip

We here discuss what can be gained by having boosted Higgses, produced in the configuration of the right panel of fig.~\ref{fig:beams}.
The main new qualitative feature is that most final states from Higgs decay are in a small forward cone --- a region near to the beam pipe
that is particularly problematic at a muon collider.
This would affect generic measurements allowing, for
example, to independently measure the Higgs mass from the
angular distribution of its decay products.

\smallskip

The focusing feature become advantageous when performing specific searches for particles $\phi$ 
(for example hypothetical scalars $\phi$ produced in $h \to \phi \bar \phi$ decays and dubbed `dark Higgs bosons')
that are long-lived and interact weakly with SM particles in the crossed surrounding material, 
producing a visible final state away from the collision point when they decay or scatter with the material.
The advantage arises because a detector away from the collision
point can only cover a small angular size $\Omega$ around the collision.
It is thereby convenient that the Higgs boost concentrates all $\phi$ arising from its decay in a 
small cone with $\Omega \simeq \pi \theta_\phi^2$,
where $\theta_\phi$ is the $\phi$ polar angle with respect to the Higgs direction in the laboratory frame.
Placing the detector in the decay cone allows to gain sensitivity.
The gain factor is limited when considering
a detector sensitive to $\phi$ particles that decay or scatter only within its {volume},
as its sensitivity is maximal when placed at a distance from the collision point comparable to the size of the detector itself,
even if the boosted $\phi$ life-time is much longer.
We will consider a cylindrical detector placed at 70 m distance, with 50 m length and with 3 m radius.

\smallskip

\begin{figure}[t]
\begin{center}
$$\includegraphics[width=0.48\textwidth]{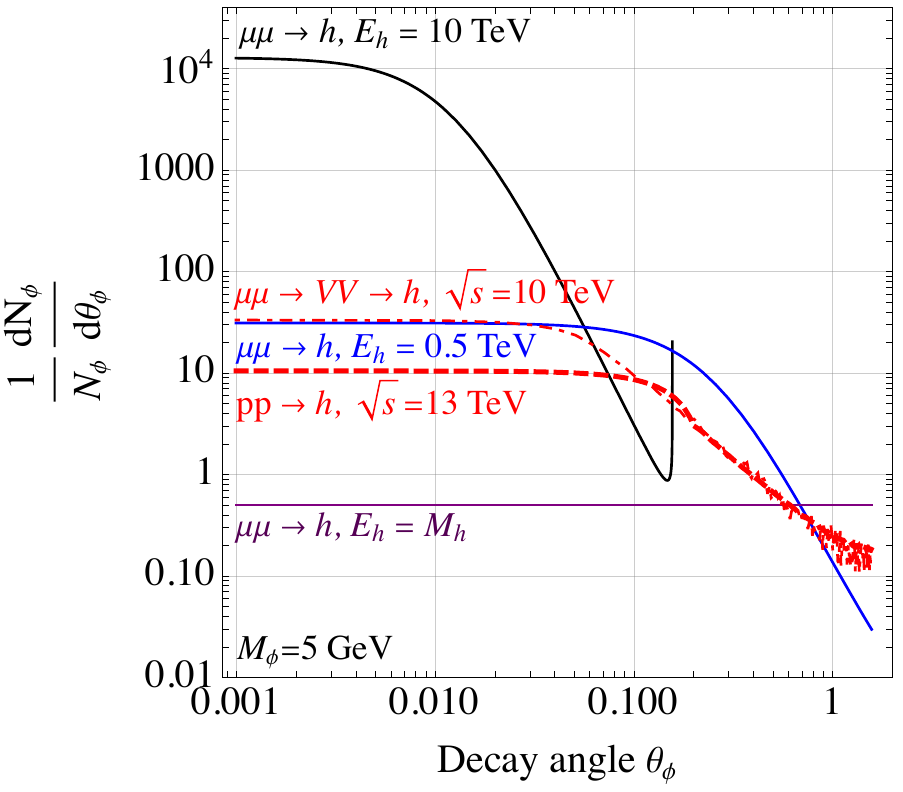}\qquad
\includegraphics[width=0.45\textwidth]{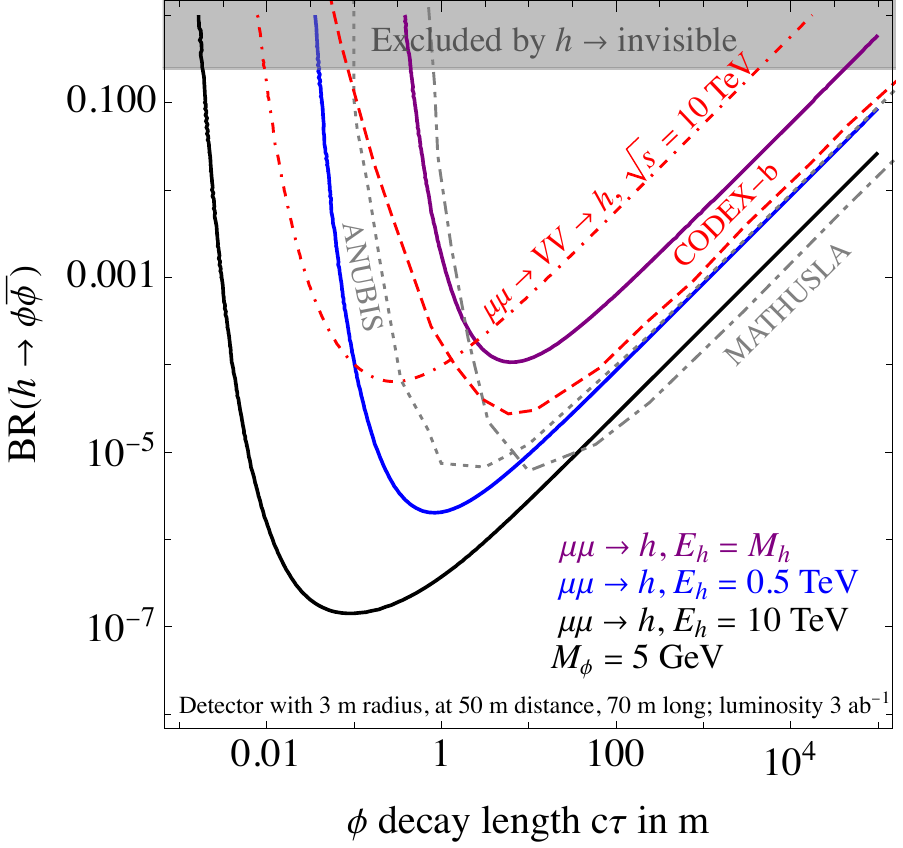}
$$
\caption{\label{fig:dNtheta}
{\bfseries Left}: Angular distribution of $\phi$ particles per one $h\to \phi\bar \phi$ decay.
We consider the Higgs $h$ produced as $\mu^-\mu^+\to h$ at a boosted muon collider,
as $\mu^-\mu^+\to VV \to h$ at a symmetric muon collider,
as $pp\to h$ at the LHC collider with $\sqrt{s}=13\TeV$.
So a boosted muon collider 
can produce much higher fluxes in the forward direction $\theta_\phi\ll 1$.
{\bfseries Right}: the resulting sensitivities, taking into account the
cross sections $\sigma(\mu^-\mu^+\to h)\approx 22\,{\rm pb}$,
$\sigma(\mu^-\mu^+\to VV\to h)\approx 0.85\pb$,
$\sigma(pp\to h)\approx50\pb$, 
and considering a far-away detector with $\sim 2000\,{\rm m}^3$ volume, such as {\sc Codex-b}~\cite{2203.07316}.
At given luminosity, a boosted $\mu^-\mu^+\to h$ collider offers a much higher sensitivity
than a symmetric muon collider and than LHC.
For comparison we also show the sensitivity of larger
far-away detectors {\sc Anubis} (about 6 times larger volume~\cite{1909.13022}, see also~\cite{anubis_new} for a recent design proposal) and
{\sc Mathulsa} (about 130 times larger~\cite{1811.00927}). 
} 
\end{center}
\end{figure}

We now compute the sensitivity on long-lived particles from Higgs decay
showing that it 
can be a few orders of magnitude better than the sensitivity of a similar detector
with a similar luminosity at the LHC $pp$ collider, as well as than at a symmetric muon collider.
We consider the specific example of a $h \to \phi \bar \phi$ decay, 
where $\phi$ can be a SM particle or an hypothetical new particle
with mass $M_\phi< M_h/2$.
The $\phi$ particles tend to be concentrated along the Higgs direction.
The Higgs has spin 0, so $h \to \phi\bar \phi$ is isotropic in the Higgs rest frame.
The angular distribution in the laboratory frame in terms of 
$\epsilon  = 2M_\phi/M_h$ and $X = \cos^2\theta_\phi ( \gamma_h^2-1) - \gamma_h^2$ is
\beq \label{eq:dNdc}
\frac{dN_\phi}{d\cos\theta_\phi}= 
\frac{X (1+ \epsilon ^2 \gamma _h^2)+ 2 \gamma _h \left[\gamma _h + \cos\theta_\phi \sqrt{(\gamma _h^2-1)(1+X \epsilon ^2}) \right]}
{2X^2 \sqrt{(1-\epsilon ^2) (1+X \epsilon ^2)}} 
\simeq  \frac{2\gamma_h^2}{(1+ \gamma_h^2\theta_\phi^2)^2} 
\eeq
normalized such that $dN_\phi/d\cos\theta_\phi = 1/2$ for $\gamma_h=1$.
The latter expression in eq.\eq{dNdc} holds in the forward limit 
$\theta_\phi  \ll 1/\gamma_h$ for large $\gamma_h\gg 1$ and, for simplicity, 
light $\phi$ particles $\epsilon \ll 1$.
It shows that the angular distribution is concentrated  in the forward direction $\theta_\phi =0$ within an angle $\theta_\phi \sim 1/\gamma_h$.
Furthermore, if $\epsilon \gamma_h >1$ (namely if $E_h  > M_h^2/2M_s$) all $\phi$ particles are within the cone $\theta_\phi <\theta_{\rm max}$, 
found by imposing $1+X\epsilon^2=0$:
\beq \label{eq:thetamax}
\tan^2\theta_{\rm max}=\frac{M_h^2-4 M_s^2}{4 \gamma_h^2 M_\phi^2 - M_h^2} \ .
\eeq
The angular distribution, shown inthe left panel of fig.\fig{dNtheta}, peaks at $\theta_\phi=0$, and also
features a narrow Jacobian peak at the maximal value $\theta_\phi=\theta_{\rm max}$ for $\sqrt s=10\;$TeV (black curve).
For comparison, the left panel of fig.\fig{dNtheta} also shows the flatter analogous distribution
that arises from $\mu^-\mu^+\to VV\to h$ vector-boson-fusion production at a symmetric muon collider with $s\gg M_h^2$ (dot-dashed red),
and from $pp\to h$ production at the LHC collider, considering the $g g \to h$ dominant production mode (dashed red).

The resulting sensitivity is shown in the right panel of fig.\fig{dNtheta}, as limits on the Higgs exotic branching ratio (BR) versus the proper $\phi$ decay length $c\tau$. We assume the reconstruction efficiency of the $\phi$ decay products to be 100\% and we work in the zero background hypothesis, as usually done for these types of studies. We consider a
far-away detector with $\approx 2000\,{\rm m}^3$ volume,
comparable to the {\sc Codex-b} detector being discussed for the LHC~\cite{2203.07316}. Our results are illustrated in black, blue and purple for $E_h=10\;$TeV, 500\;GeV and $M_h$ respectively.
LHC and a symmetric muon collider (dot-dashed red) offer similar sensitivity 
at similar luminosity, given that the cross sections are similar.
On the other hand, boosted Higgs production at a muon collider allows to improve the sensitivity by a few orders
of magnitude, outperforming even larger detectors such as {\sc Anubis}~\cite{1909.13022} and
{\sc Mathulsa}~\cite{1811.00927}
being discussed for the LHC.
Furthermore, the boost would allow to observe events with both particles $\phi$ within the forward detector,
providing additional information, such as the differential time-of-flight.

Fig.\fig{dNtheta} right assumes $M_\phi=5\GeV \ll M_h$. An even larger sensitivity enhancement 
arises if instead $M_\phi$ is  just a little below $M_h/2$.
In such a case $\theta_{\rm max}$ in eq.\eq{thetamax} gets smaller, 
meaning that all particles $\phi$ get concentrated in a smaller cone at a boosted $\mu$ collider.
A longer thinner detector with the same volume would allow to exploit this feature,
offering further enhanced sensitivity.

\section{Resonant $\mu^-\mu^+ \to Z$ production}\label{Z}
Similar signals to those discussed in section \ref{h} arise substituting the Higgs $h$ with the $Z$ boson
as the resonantly produced particle that decays into new long-lived states.
We focus on the differences.
The peak cross section is higher,
\beq \label{eq:sigmaZ}\sigma(\mu^-\mu^+\to Z) = 3~4\pi \, {\rm BR}(Z\to \mu^-\mu^+)/M_Z^2\approx 60\times10^3\pb ,\eeq
in view of ${\rm BR}(Z\to \mu^-\mu^+)\approx 0.0337$.
Furthermore, as the $Z$ has a larger width
$\Gamma_Z \approx 2.49\GeV \gg \Gamma_h$, 
on-shell $Z$ production is easily achieved,
and losses due to initial state radiation can be neglected.
The factor 3 in eq.\eq{sigmaZ} arises because the $Z$ boson has spin 1.
For the same reason, $Z$ decays need not being isotropic in the $Z$ rest frame.
We thereby consider specific new-physics models, where the three key phenomenological parameters
(mass and decay length of the new long-lived particle, and $Z$-boson branching ratio into the
new particle) are computed in terms of model parameters.

\smallskip

\begin{figure}[t]
\begin{center}
$$\includegraphics[width=0.48\textwidth]{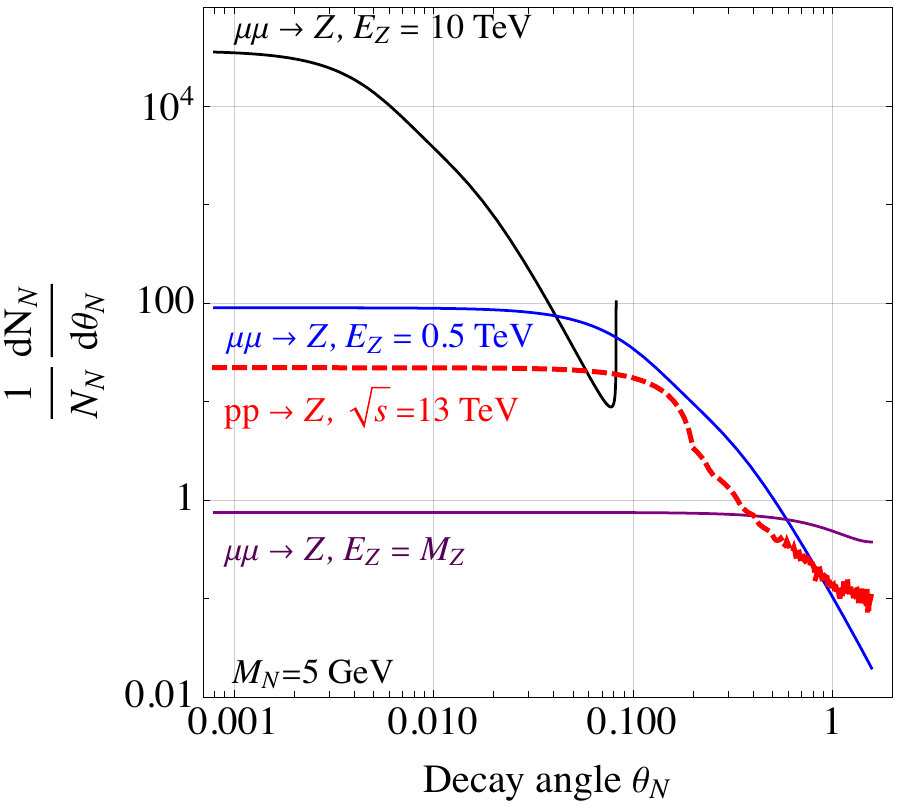}\qquad
\includegraphics[width=0.46\textwidth]{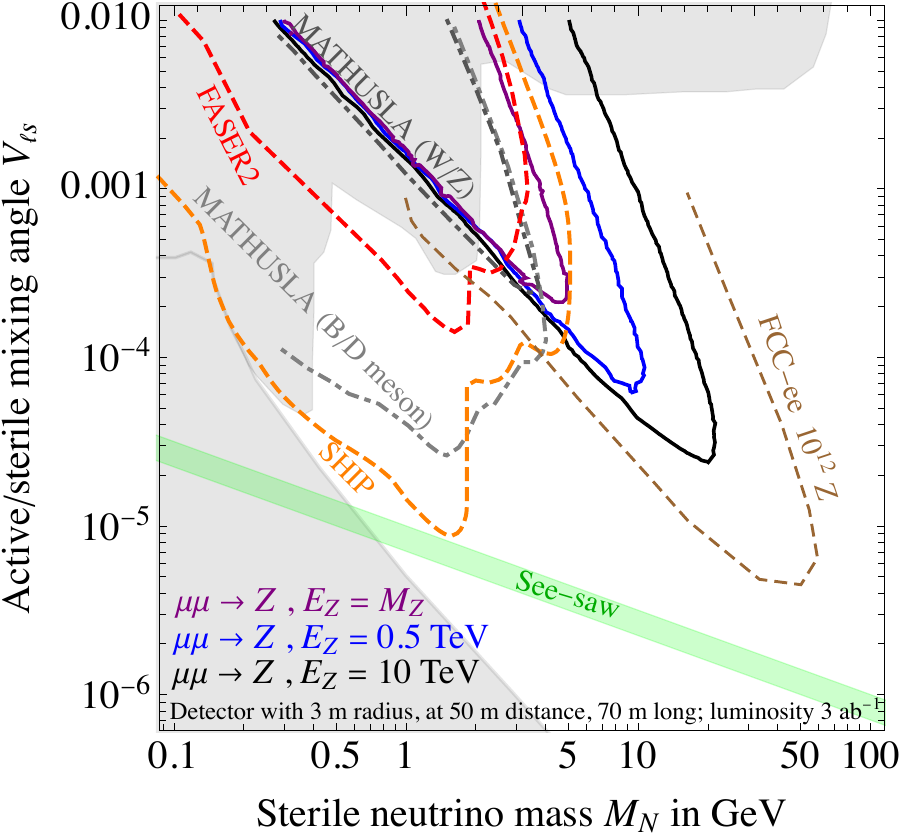}
$$
\caption{\label{fig:dNthetaZ}
{\bfseries Left}: Angular distribution of $N$ particles per one $Z\to N\nu$ decay.
We consider the $Z$ produced as $\mu^-\mu^+\to Z$ at a boosted muon collider, as $pp\to Z$ at the LHC collider with $\sqrt{s}=13\TeV$.
{\bfseries Right}: the resulting sensitivities, taking into account the
cross sections $\sigma(\mu^-\mu^+\to Z)\approx 6\times 10^4\,{\rm pb}$ and considering the same experimental setup of fig.~\ref{fig:dNtheta}.
For comparison we also show the sensitivity of other proposed experiments as well as the existing bounds from terrestrial analyses and astrophysics (in gray), all taken from~\cite{1504.04855}.
} 
\end{center}
\end{figure}

A plausible theory where the $Z$ boson decays into a long-lived particle
is the SM extended with a fermion singlet $N$ with mass $M_N$, dubbed 
`right-handed' or `sterile neutrino' because it can have the Yukawa couplings
$y_\ell \, L_\ell NH$ to left-handed leptons $L_\ell$ and to the Higgs doublet $H$.
Once $H$ acquires its vacuum expectation value $v$,
this coupling produces a mass mixing  $V_{\ell s}\simeq y_\ell v/M_N \ll 1$ with SM neutrinos,
and a contribution $(m_\nu)_{\ell \ell'} =V_{\ell s} V_{\ell's}  M_N$ to their mass matrix,
motivating small values of $V_{\ell s}$ that result in a long-lived $N$.
The main $N$ decay rates are~\cite{0705.1729,0901.3589}
\beq\Gamma(N\to  \ell \bar \ell' \nu_{\ell '},\ell q\bar q)  \sim \frac{G_{\rm F}^2 M^5_N}{96\pi^3} |V_{\ell s}|^2,\qquad
\Gamma(N\to \pi^0\nu_\ell)  \sim \frac{G_{\rm F}^2 M^3_N f_\pi^2}{32\pi} |V_{\ell s}|^2 . \eeq
Therefore the life-time $\tau_N = 1/\Gamma_N$ steeply scales  with $M_N$.
On the other hand, the $Z$-boson decay rate dominantly depends on the mixing~\cite{Dittmar:1989yg}
\beq 
{\rm BR}(Z\to \nu N)\simeq {\rm BR}(Z\to \nu \nu) \frac{2}{3}\sum_\ell |V_{\ell s}|^2
\left(1 - \frac{M_N^2}{M_Z^2}\right)^2\left(1+\frac{M_N^2}{2M_Z^2}\right)
\eeq
where ${\rm BR}(Z\to \nu\bar\nu)\approx 20.0\%$. 
The angular distribution of the sterile neutrino $N$ produced in $Z$ decays is shown in the left panel of fig.~\ref{fig:dNthetaZ}. 
This plot is similar to fig.\fig{dNtheta}, left panel, for Higgs decays, except that we have here omitted
vector-boson-fusion $\mu^-\mu^+\to Z$ production at a symmetric muon collider,
in view of its small cross section, about 4 pb at $\sqrt s = 10$ TeV. 

Constraints and sensitivities are presented in the right panel of fig.\fig{dNthetaZ}.
We consider a far away detector with the same geometry (volume and distance) as in  fig.\fig{dNtheta}, right panel, and we adopt the same assumptions on reconstruction efficiency and background.
While fig.\fig{dNtheta} was made  in the phenomenological plane $(c\tau, {\rm BR})$,
fig.\fig{dNthetaZ} uses the model parameters ($M_N, |V_{\ell s}|^2$),
possibly restricted along or below the green band where $m_\nu$ acquires the observed values
(assuming either normal or inverted neutrino mass hierarchy).
Using model parameters allows to consider a variety of different experiments: colliders, fixed target, meson decays.
The various existing bounds are plotted as dark shadows, from~\cite{1504.04855}.
The sensitivities of possible experiments are plotted as dashed curves:
{\sc Faser2}~\cite{1811.12522} and the  larger proposed detector  {\sc Mathusla} mostly look at decays of mesons produced at the LHC,
while the proposed {\sc SHiP} experiment would look at mesons produced by fixed-target collisions~\cite{1504.04855, 2305.13383}.
We also show the reach of  {\sc Mathusla}~\cite{1806.07396} when looking at $Z,W$ decays.
A boosted muon collider offers a comparable reach on the mixing angle, 
but in a different region with larger mass $M_N$, 
since sterile neutrinos arise from boosted $Z$ decays.
Other proposed detectors such as {\sc Anubis} and {\sc Codex-b} offer a comparable or weaker reach~\cite{1911.06576,2001.04750}, and we do not show them for clarity. The same consideration applies for similar far-detector at the proposed FCC-hh experiment, see {\it e.g.}~\cite{2204.01622}.

Furthermore, $Z$ bosons can also be resonantly produced at an $e^- e^+$ collider:
a future circular $e^- e^+$ collider with 100 km length could perhaps
produce $10^{12}$ $Z$ bosons, corresponding to a ${\cal L}\sim 20/\ab$ integrated luminosity~\cite{CEPC,FCCee},
larger that what assumed for a muon collider.
Such a  100 km $e^- e^+$ collider,  with a full $4\pi$ detector that captures all $Z\to \nu N$ decays (in the mass range where $\tau_N$ is not too large),
offers higher sensitivity that the assumed boosted muon collider, with small-angle detector and lower luminosity.
None of these proposals reaches the band denoted as `see-saw', where the neutrino masses mediated by the sterile
neutrino match the measured neutrino masses.

\smallskip

If multiple quasi-degenerate sterile neutrinos exist, a boost could help studying their oscillations, similarly to
what done with mesons at $e^-e^+$ asymmetric colliders.

Similar results hold for different models, such as a new vector $V$ produced as $Z\to \gamma V$.

\section{Non resonant production of long-lived particles}\label{NR}
The previous sections assumed that the SM $h,Z$ particles act
as mediators between muons and new long-lived states.
This made convenient having a collider running at the resonance $\sqrt{s}=M_{h,Z}$ for $h,Z$ production,
rather than at the maximal possible $\sqrt{s}$.

More in general, mediators could be new unknown particles exchanged in the $s$ or $t$ channel:
in such a case running at reduced $\sqrt{s} < 2E_{\rm beam}$ around the unknown mediator masses would similarly
enhance the sensitivity to long-lived particles.

If mediators are heavier than the collider energy, physics gets approximated via effective operators.
In this idealised limit the cross sections for producing new long-lived light particles
can grow with the collision energy,
reducing the advantage of running a collider at  $\sqrt{s}$ below the maximal $\sqrt{s} =2 E_{\rm beam}$.
Let us consider two examples.
\begin{itemize}
\item A dimension 5 effective operator, such as a sterile neutrino $N$ coupled to photons via a magnetic dipole moment operator
$F_{\mu\nu}(\bar\nu \gamma_{\mu\nu} N)/\Lambda$.
The resulting cross section $\sigma(\mu^-\mu^+\to \gamma^* \to \nu N)\sim e^2/\Lambda^2$
is energy-independent~\cite{1909.04665}, so considerations similar to previous sections apply.

\item A dimension 6 effective operator, such as a sterile neutrino $N$ coupled to SM fermions via $(\bar\mu \gamma_\alpha \mu) (\bar\nu \gamma^\alpha N)/\Lambda^2$ operators.
In such a case the $\gamma^2 = E^2_{\rm beam}/s$ enhancement of the $N$ flux in the boosted direction
gets compensated
by the energy dependence of the production cross section
$\sigma(\mu^-\mu^+\to \nu N) \sim s/\Lambda^4$
(and possibly by the boosted decay length, if it exceeds the detector size).
As a result a boosted collider would be  convenient only if it could deliver an enhanced luminosity
compared to a symmetric collider~\cite{2105.13851}.
\end{itemize}
Rather than studying in the detail all possibilities, we conclude with a panoramic summary of the main points.

\section{Conclusions}\label{concl}
We explored the possibility of running a collider
in `boosted' configuration,
with beam energies higher than the collision energy $\sqrt{s}$, thereby 
producing particles boosted by a factor $\gamma \approx E_{\rm beam}/\sqrt{s}$.
We considered producing the heavier SM particles, such as the Higgs and the $Z$ boson.
Their boosted production is not done at $e^- e^+$ colliders because it implies a big luminosity loss at fixed consumed power.
The situation is different at a $\mu^-\mu^+$ collider, where
the luminosity of head-on symmetric collisions is expected to grow proportionally to $s$.
In section~\ref{boost} we estimated that achieving a boost $\gamma$ affects the luminosity as
\beq \label{eq:L/L}
\frac{{\cal L}_{\rm boosted}}{{\cal L}_{\rm symmetric}}\sim\left\{
\begin{array}{ll}
1/\gamma^{8} & \hbox{$e^-e^+$ head-on collisions with asymmetric beam energies,}\\
1/\gamma^2 & \hbox{$\mu^-\mu^+$ head-on collisions with asymmetric beam energies,}\\
\gamma & \hbox{$\mu^-\mu^+$ oblique collision with same beam energies}.
\end{array}\right.
\eeq
Therefore the simplest head-on geometry gives a mild luminosity loss at a $\mu^-\mu^+$ collider,
while a luminosity enhancement could potentially arise performing oblique collisions,
if focused bunches can be tilted by an appropriate large angle.
While small-angle tilting is considered possible~\cite{Calaga:2018ipx}, 
a large angle would need a dedicated beam optics:
whether this is possible or not is a key aspect beyond the scope of this work, which we do not investigate further. We instead concentrated on identifying interesting physics cases that can take advantage of these non-standard collider options.

\medskip

Next, we explored what can be achieved by having boosted particles.
The boost significantly helps one  search for specific new physics: long-lived new particles
that can be best detected in a far-away detector.
The reason is that the far-away detector can only cover a relatively small solid
angle $\Omega\ll 4\pi$ around the interaction point, and thereby placing it along the boost direction
enhances the signal rate by a factor $\gamma^2$,
possibly reduced down to a factor $\gamma$ if the particles are so much long-lived that
boosted decays happen beyond the detector.
Thereby the number of detected events, that controls the sensitivity of the search, scales as
\beq \frac{N_{\rm boosted}}{N_{\rm symmetric}} \sim  \gamma^{1-2}
 \frac{{\cal L}_{\rm boosted} }{{\cal L}_{\rm symmetric}}
 \frac{\sigma (E^2)}{\sigma(s)} 
 \ .
\eeq
The production cross sections can scale in different ways with energy, depending on the model. 
The main possibilities are:
\beq  \frac{\sigma (E^2)}{\sigma(s)} \sim\left\{
\begin{array}{ll}
\gamma^2 & \hbox{Decays of $Z,h$ mediators, produced resonantly at $s=M_{h,Z}^2$}\\
1 & \hbox{Dimension 5 effective operators}.\\
1/\gamma^2 & \hbox{Dimension 6 effective operators.}
\end{array}\right.
\eeq
The most optimistic win-win-win situation would provide 
a gain $N_{\rm boosted}/N_{\rm symmetric}\sim\gamma^5$
in the case where resonantly produced boosted Higgs bosons decay into mildly long-lived new particles,
assuming that oblique collisions can enhance the luminosity of a $\mu^-\mu^+$ collider as in eq.\eq{L/L}. 
The boost factor could be $\gamma \sim E/M_h \sim 10$ or more.
Fig.\fig{dNtheta} shows that boosting the Higgs can provide higher sensitivity than other proposals,
having assumed the same luminosity and a relatively small far detector, such as {\sc Codex-b}.

Analogously, section~\ref{Z} studies $Z$ decays into long-lived new particles and fig.\fig{dNthetaZ} shows 
that a boosted $\mu^-\mu^+\to Z$ collider can compete with other proposals.

We conclude by stressing that, in order to assess the robustness of our results, various experimental effects must be investigated further. These includes 
 beam-induced background effects, that might have a significant impact in the forward-direction even if the detector is positioned far-away from the interaction point, together with beam-energy spread effects that might deplete the resonant cross-section for $h,Z$ production.

\small

\subsubsection*{Acknowledgments}
This work was supported by the MIUR grant PRIN 2017L5W2PT.
We thank Rama Calaga, Stephane Fartoukh, Roberto Franceschini
Filip Moortgat, Eugenio Paoloni, Arsenii Titov and Andrea Wulzer for useful discussions on collider aspects.

\end{document}